\documentstyle[seceq,preprint,epsfig]{ptptex}

\preprintnumber[3cm]{%
SNUTP-03-028}
\title{Color Superconducting Gap in Schwinger-Dyson Equation
and Nonlocal Gauge Fixing}

\author{
Chaejun {\sc Song}
}

\inst{
School of Physics and CTP, Seoul National University, Seoul 
151-747, Korea}

\abst{
We solve Schwinger-Dyson (SD) equation for the color superconducting (CS)
gap in nonlocal gauge which justifies the ladder approximation in subleading
order calculation. The gap prefactor increases about
1.6 times to the leading order gap determined in Landau
gauge.
}

\begin{document}

\maketitle

\section{Introduction}
High density quark matter is expected to be in CS phase.
Many approach CS phase from the asymptotic densities since 
perturbative QCD can be used.
Son showed that the gap increases
due to the long-range magnetic interactions as density increases~\cite{son}. 
And many works are done in order to determine the size of the gap 
in leading and subleading 
order~\cite{hong1, hong2,leading, subleading}.

In the leading order analysis of SD 
equation without vertex and wavefunction renormalization
corrections, the gap prefactor is found to be 
gauge-dependent~\cite{hong1,hong2}
though gauge-dependence is negligible at extremely high densities.
It's not so surprise.  
The solutions of the SD equation with the bare vertex  
can be strongly gauge dependent
even in QED without any matter~\cite{nonoyama} since
the simple ladder approximation, where the full vertex function
is approximated by the bare one, is not consistent with 
Ward-Takahashi (WT) identity. 

We solve the SD equation for the CS 
gap, respecting gauge invariance as much as possible. We try to 
find the nonlocal gauge~\cite{georgi}, which makes the simple
ladder approximation consistent with WT identity in the
gap calculations to subleading order. We consider just color-flavor-locked
CS phase here for simplicity but there is no difference in two flvor
CS phase.  

\section{Nonlocal gauge}
The inverse propagator of the Nambu-Gorkov quark field
$\Psi(x)\equiv(\psi(x),\psi_c(x))^T$
is \begin{equation}
S^{-1}(p)= -i\begin{pmatrix}
a(p)\left[(p_0+\mu)\gamma^0+b(p)\not\!\vec p\,\right] &
       -\Delta(p) \cr
-\gamma^0\Delta^{\dagger}(p)\gamma^0&
a(p)\left[(p_0-\mu)\gamma^0+b(p)\not\!\vec p\,\right]\cr
\end{pmatrix},
\label{inverse}
\end{equation}
with wavefunction renormalization constants $a(p), b(p)$ and
the gap $\Delta$. 

We want to find the nonlocal gauge where
the wavefunction renormalization constants are unchanged
by renormalization corrections, i.e.,
$a(p)=b(p)=1$. In QED,
such a nonlocal gauge enables us to use the bare vertex without
breaking WT identity in all orders~\cite{georgi}. 
In contrast, WT identity of QCD (Slavonv-Taylor
identity) is more complicated and nonlocal gauge does not
guarantee that we can keep the bare vertex with gauge symmetry
in all orders.
However, we can keep the bare vertex $\gamma^\mu$ with WT identity
in high density (so weak coupling) limit. 
\begin{figure}
\epsfxsize=5in
\centerline{\epsffile{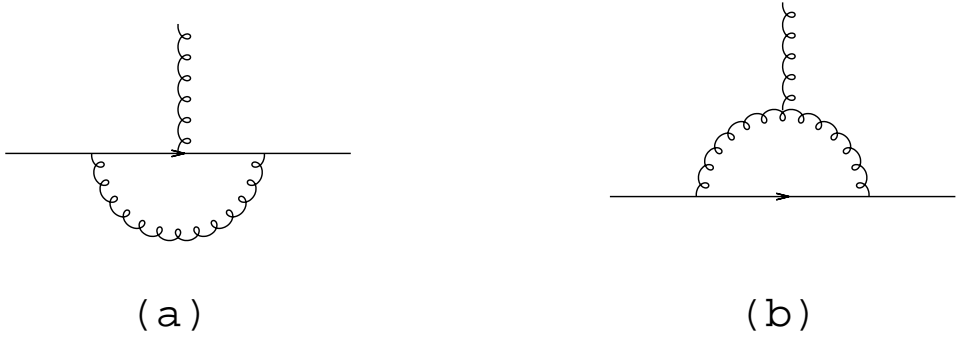}}
\caption{The solid line denotes quarks and the curly lines
gluons.}
 \label{vertex}
\end{figure}

The QED-like diagram (Fig.~\ref{vertex}a) gives the relations
\begin{eqnarray}
(p-p^{\prime})_{\mu}\Lambda^{(a)\mu}(p,p^{\prime})
=a(p)\left[(p_0+\mu)\gamma^0+b(p)\not\!\vec p\right]-
a(p^{\prime})\left[(p^{\prime}_0+\mu)\gamma^0+b(p^{\prime})\not\!
\vec p^{\prime}\right],
\nonumber\end{eqnarray}
where we suppress the color indices. We see that our nonlocal 
gauge where $a(p)=b(p)=1$ makes $\Lambda^{(a)\mu}=\gamma^{\mu}$.
The trigluonic diagram (Fig.~\ref{vertex}b) contributes
in high density effective theory~\cite{hong1} as
\begin{eqnarray}
I_{\mu}^a&\simeq&\,g_s^3
\int_l{l\cdot V\over l_{\parallel}^2+\Delta^2}{
\gamma_0f^{abc}T^bT^c\left[\left(c_1l_{\parallel}^i+
c_2l_{\perp}^i+c_3p^i+c_4p^{\prime i}\right)
g_{\mu i}+c_5\left(2l-p-p^{\prime}\right)_{\mu}\right]\over
\left[\left|\vec l-\vec p\right|^2+\pi M^2\left|l_0-p_0\right|/
\left(2\left|\vec l-\vec p\right|\right)\right]\left[
\left|\vec l-\vec {p^{\prime}}\right|^2+\pi M^2\left|l_0-p^{\prime}_0\right|/
\left(2\left|\vec l-\vec {p^{\prime}}\right|\right)
\right]}\nonumber\\
&\sim & i\,g_s^3\gamma_0V_{\mu}T^a\left({\Delta\over M}\right)^2\ln\left(
{\Delta/\mu}\right),\nonumber
\end{eqnarray}
where $c_1$ and $c_5$ are $1+O\left(\Delta^2/l_{\perp}^2\right)$ while
all other $c_i$'s are $O\left(\Delta^2/l_{\perp}^2\right)$.
The screening mass $M$ is given as $g_s\mu\sqrt{N_f}/(2\pi)$ for
$N_f$ light quarks in the hard-dense-loop (HDL) approximation.
The trigluonic contribution is suppressed, compared with QED-like diagram
contribution. We can use the simple ladder approximation for
the SD gap equation to subleading order
after determining the proper nonloacal gauge.

After projecting out the antiquarks by the
on-shell projectors $\Lambda_p^\pm\equiv
\frac12\  (1\pm\frac{\gamma^0\vec{\gamma}\cdot\vec{p}}{|\vec{p}|}
)$, the equation~\cite{hong2} for the local gauge where $a(p)=b(p)=1$
is given as
\begin{eqnarray}
-\frac32\pi\alpha_s\int\frac{d^4q}{(2\pi )^4}D_{\mu\nu}(q-p)
\frac{q_0+|\vec{q}|-\mu}{q_0^2
-\{|\vec{q}|-\mu\}^2-\Delta^2}
{\rm tr}\, \ [\gamma^\mu\gamma^0\Lambda^-_q\gamma^\nu\gamma^0
\Lambda^\pm_p]
=0
\label{waveq}\end{eqnarray}
where \begin{equation}
D_{\mu\nu}(k)=D^{(1)}\, O_{\mu\nu}^{(1)} 
+D^{(2)}\, O_{\mu\nu}^{(2)}+
D^{(3)}\, O_{\mu\nu}^{(3)}\,,
\end{equation}
where in the weak coupling limit, $|k_0|\ll|\vec k|$,
\begin{eqnarray}
D^{(1)}={\left|\vec k\right|\over \left|\vec k\right|^3+M_0^2\Delta+\pi M^2
\left|k_4\right|/2}\,,\quad
D^{(2)}={1\over k_4^2+{\vec k}^2+2M^2}\,,\quad
D^{(3)}={\xi \over k_4^2+{\vec k}^2},
\nonumber\end{eqnarray}
where $M_0$ is Higgs-like gluon mass
$\sim g_s\mu /(2\pi)$~\cite{hong2}.
The polarization tensors are defined as
\begin{eqnarray}
O^{(1)}=P^{\perp}+{(u\cdot k)^2\over (u\cdot k)^2-k^2}P^u,
O^{(2)}=P^{\perp}-O^{(1)},\quad O^{(3)}=P^{\parallel},
\nonumber\end{eqnarray}
where $u_{\mu}=(1,\vec 0)$ and
\begin{eqnarray}
P^{\perp}_{\mu\nu}=g_{\mu\nu}-{k_{\mu}k_{\nu}\over k^2},\quad
P^{\parallel}_{\mu\nu}={k_{\mu}k_{\nu}\over k^2},\nonumber\\
P^u_{\mu\nu}={k_{\mu}k_{\nu}\over k^2}-
{k_{\mu}u_{\nu}+u_{\mu}k_{\nu}\over (u\cdot k)}
+{u_{\mu}u_{\nu}\over (u\cdot k)^2} k^2.
\nonumber\end{eqnarray}
Now, we assume the (nonlocal) gauge fixing parameter $\xi(k)$
has only temporal dependence, that
is, $\xi(k)\simeq\xi(k_4)$. After angluar integration
with $|\vec{p}|,\  |\vec{q}|\sim\mu$, (\ref{waveq})
becomes;
\begin{equation}
I^{(1)}+I^{(2)}+I^{(3)}=0
\label{eq2}
\end{equation}
where
\begin{eqnarray}
I^{(1)}
\approx \frac{2}{3}\,\ln \left[\frac{(2\mu)^3}{
M_0^2\Delta +\pi M^2|p_4-q_4|/4} \right],
I^{(2)}
\approx 1, 
I^{(3)}
\approx-\xi\, \ln\left[ \frac{ (2\mu)^2}{
|p_4-q_4|^2}\right].
\nonumber
\end{eqnarray}
We find the solution of  Eq.~(\ref{eq2});~\cite{hlmss}
\begin{equation}
\xi\approx \frac{\frac{2}{3} \ln \frac{(2\mu)^3}{ M_0^2\Delta +
\pi M^2|p_4-q_4|/2}}{\ln \frac{ (2\mu)^2}{|p_4-q_4|^2}}.
\label{gauge}
\end{equation}

\section{Gap equation}
With our nonlocal gauge (\ref{gauge}) the SD equation 
for the gap is~\cite{hong2}
\begin{eqnarray}
\Delta (p_4)
&=&\frac23\pi\alpha_s\int\frac{d^4q}{(2\pi)^4} D_{\mu\nu}^\xi (q-p)
\Delta (q)
\frac{{\rm tr}\,  [\gamma^\mu\Lambda^+_q\gamma^\nu\Lambda^-_p]}{q_0^2
-\{|\vec{q}|-\mu\}^2-\Delta^2}\\
&\sim& \frac{2\alpha_s}{9\pi}\int_0^{p_4} dq_4
\frac{\Delta (q_4)}{\sqrt{q_4^2+|\Delta (q_4)|^2}}
\log\frac{\Lambda (p_4)}{p_4}
+
\frac{2\alpha_s}{9\pi}\int^\Lambda_{p_4} dq_4
\frac{\Delta (q_4)}{\sqrt{q_4^2+|\Delta (q_4)|^2}}
\log\frac{\Lambda (q_4)}{q_4}\nonumber\end{eqnarray}
where $\Lambda \equiv e^{3\xi/2}(2\mu)^6/(\sqrt2\pi M^5)$.
We can replace the integral equation by the differential equation;
\begin{equation}
p\frac{d^2\Delta}{dp^2}+\frac{d\Delta}{dp}
+\frac{2\alpha_s}{9\pi}\frac{\Delta (p)}{
\sqrt{p^2+|\Delta|^2}}=0.
\nonumber\end{equation}
Here we neglect the contributions related with $\frac{d\Lambda}{dp},
\frac{d^2\Lambda}{dp^2}$ since they are suppressed by the condition
$|p_4|\ll M_0, M$. 

In IR region ($p\ll |\Delta|$) with $p\frac{d\Delta}{dp}\mid_{p=0}=0$,
$\Delta (p)=\Delta_0 J_0 (\sqrt{\frac{8\alpha_s p}{9\pi\Delta_0}})$. 
And in UV region ($p\ll |\Delta|$) with
$p\frac{d\Delta}{dp}\mid_{p=0}=0$, 
$\Delta (p)=B\sin (\sqrt{\frac{2\alpha_s}{9\pi}}\log\frac{\Lambda}{p})$.
We can determine $\Delta_0$ (and $B$) by matching two solutions 
at $p=\Delta_0$. In the weak coupling limit~\cite{hong2}, 
\begin{equation}
\Delta_0\approx \frac{2^7\pi^4}{g^5}\mu e^{1+\frac32\xi}
e^{-\frac{3\pi^2}{\sqrt2 g}}.
\end{equation}
Note that $\xi \sim 1/3$ in Eq. (\ref{gauge}) with $|p_4-q_4|\sim\Delta$.
So the vertex and wavefunction corrections increase the gap prefactor
$e^{1/2}$ times larger than the leading order result at Landau gauge 
($\xi =0$)~\cite{leading}.
\section{conclusion}
We solve the SD equation for the CS gap after determining the nonlocal gauge
where wavefunction renormalization corrections vanish. Our nonlocal
gauge makes us be able to keep
WT identity for the gap calculations to subleading order
and to make the bare vertex (ladder) approximation available.
The nonlocal gauge is $\sim 1/3$ near on-shell of the gap
and increases the leading-order gap at Landau gauge by 2/3.

%
%
\section*{Acknowledgements}
Most of the contents are done with D.~K.~Hong, T.~Lee, D.~-P.~Min and
D.~Seo and are published in Ref.~8.
We would like to thank M.~Alford, M.~Harada, T.~Hatsuda and K.~Yamawaki
for their comments and the organizing members for their hospitality
in Nara. 
We were supported by KRF 2001-015-DP0085 and BK21
projects of Ministry of Education, Republic of Korea.

%
%


\begin{thebibliography}{99}
\bibitem{son}
D.~T.~Son, \PR{D59,1999,094019}. 
\bibitem{hong1} D.~K.~Hong, \PL{B473,2000,118}; \NP{B582,2000,451}.
\bibitem{hong2}
D.~K.~Hong, V.~A.~Miransky, I.~A.~Shovkovy and L.~C.~R.~Wijewardhana,
\PR{D61,2000,056001}; {\it Erratum} \PR{D62,2000,059903}. 
\bibitem{leading} T.~Sch\"afer and F.~Wilczek, \PR{D60,1999,114033};
P.~Pisarski and D.~Rischke, \PR{D61,2000,074017}.
\bibitem{subleading} W.~E.~Brown, J.~T.~Liu and H.~C. Ren, 
\PR{D61,2000,114012}; C. Manuel, \PR{D62,2000,114008};
S.~R.~Beane, P.~F.~Bedaque and M.~J.~Savage, \NP{A688,2001,931};
Q.~Wang and D.~H.~Rischke, \PR{D65,2002,117502}; T.~Sch\"afer,
\NP{A728,2003,251}.
\bibitem{nonoyama}
T.~Nonoyama and M.~Tanabashi, \PTP{81,1989,209}.
\bibitem{georgi} H.~Georgi, E.~H.~Simmons and A.~G.~Cohen, 
\PL{B236,1990,183}.
\bibitem{hlmss} D.~K.~Hong, T.~Lee, D.~P.~Min, D.~Seo and C.~Song,
\PL{B565,2003,153}.
\end{thebibliography}
\end{document}